\documentclass[aps,prb,twocolumn,groupedaddress,showpacs,floatfix,altaffilletter,longbibliography]{revtex4-2}
\usepackage{graphicx}
\usepackage{grffile}
\usepackage{amsmath}
\usepackage{amssymb}
\usepackage{bm}
\usepackage{color}
\usepackage[dvipsnames]{xcolor}
\usepackage{amsmath,amssymb,amsfonts}
\usepackage{epsfig}
\usepackage{times}
\usepackage[colorlinks,bookmarks=false,citecolor=blue,linkcolor=red,urlcolor=blue]{hyperref}
\usepackage{gensymb}
\usepackage[toc,page]{appendix}
\usepackage{float}

\newcommand{\fref}[1]{Fig.~\ref{#1}}
\newcommand{\eref}[1]{Eq.~(\ref{#1})} 

\newcommand{\appref}[1]{Appendix~\ref{#1}}

\begin{document}

\title{Disorder effects on hot spots in electron-doped cuprates}

\author{C. Gauvin-Ndiaye$^{1}$, P.-A. Graham$^{1}$, and A.-M.S.~Tremblay$^{1}$}
\affiliation{$^1$D{\'e}partement de Physique, Institut quantique, and RQMP Universit{\'e} de Sherbrooke, Sherbrooke, Qu{\'e}bec, Canada  J1K 2R1}
\date{\today}
\begin{abstract}
Antiferromagnetic fluctuations in two dimensions cause a decrease in spectral weight at so-called hot spots associated with the pseudogap in electron-doped cuprates. In the $2D$ Hubbard model, these hot spots occur when the Vilk criterion is satisfied, namely when the spin correlation length exceeds the thermal de Broglie wavelength. Using the two-particle self-consistent approach, here we show that this criterion may be violated near the antiferromagnetic quantum critical point when sufficient disorder is added to the model. Static disorder decreases inelastic scattering, in contradiction with Matthiessen's rule, leading to a shift in the position of the quantum critical point and a modification of the conditions for the appearance of hot spots. This opens the road to a study of the interplay between disorder, antiferromagnetic fluctuations and superconductivity in electron-doped cuprates.
\end{abstract}

\maketitle

\section{Introduction}
\label{sec:intro}
Cuprate superconductors are notorious in good part because of their normal state where a loss of Fermi-level spectral weight, so-called pseudogap, appears~\cite{Proust_Taillefer_2019}. In the case of electron-doped cuprates~\cite{Armitage:2010}, the normal state shows a loss of spectral weight in ARPES measurements at the hot spots where the Fermi surface crosses the antiferromagnetic Brillouin zone \cite{Armitage_2002,Matsui_2005,Horio_2016,He_2019}. This loss of spectral weight at hot spots, that we call an antiferromagnetic (AFM) pseudogap, has also been probed by angle-dependent magnetoresistance oscillations \cite{Helm_2010,Kartsovnik_2011}, Shubnikov-de Haas oscillations \cite{Helm_2009,Helm_2010}, optical conductivity \cite{Onose_2004, Zimmers_2005} and it has been linked to commensurate $(\pi,\pi)$-spin-density-wave fluctuations (AFM spin fluctuations for short) both experimentally \cite{Motoyama_2007,Boschini_2020} and theoretically \cite{Vilk_1997, Vilk_1997b, Kyung_2004,Schafer_2021}. The presence of AFM spin fluctuations at finite temperature in electron-doped cuprates is not surprising since there is an AFM ground state that ends at a quantum critical point (QCP)~\cite{Mandal_Sarkar_Greene_2019}. 

Theoretically, an AFM pseudogap can appear in a wide temperature range in two-dimensions ($2D$) as a precursor of long-range order that occurs only at zero temperature, as imposed by the Mermin-Wagner theorem. In the weak to intermediate interaction limit, the opening of the AFM pseudogap has been shown within the one-band Hubbard model to occur in the renormalized classical regime when the Vilk criterion is satisfied, namely when the spin correlation length $\xi_{sp}$ becomes larger than the thermal de Broglie wavelength $\xi_{th}=\hbar v_F/\pi k_B T$, where $v_F$ is the Fermi velocity \cite{Vilk_Tremblay_1995,Vilk_1997,Vilk_1997b, Schafer_2021}. Experimentally, Motoyama \textit{et al.} showed that the pseudogap temperature $T^*$ of the electron-doped cuprate Nd$_{2-x}$Ce$_x$CuO$_4$ (NCCO) does satisfy  the Vilk criterion, namely $\xi_{sp}(T^*) \simeq 2.6 \xi_{th}(T*)$, in the underdoped samples~\cite{Motoyama_2007}. However, the Vilk criterion no longer holds in optimally~\cite{Motoyama_2007} and overdoped~\cite{He_2019} samples where the AFM correlation length is small. It then becomes relevant to investigate more deeply how short-range spin fluctuations can lead to an AFM pseudogap.

One of the original mechanisms proposed to explain the failure of the Vilk criterion at optimal doping is that disorder could introduce an additional relevant length scale in electron-doped cuprates at low temperatures \cite{Motoyama_2007}. Multiple factors influence the level of disorder in these materials, notably cerium doping, that introduces impurities as well as local deformations of the lattice and a contraction of the c-axis in Nd$_{1.85}$Ce$_{0.15}$CuO$_4$ \cite{Sperandini_1998}. In its ideal form, the structure of electron-doped cuprates such as Pr$_{2-x}$Ce$_x$CuO$_4$ (PCCO) and NCCO, called $T'$,  contains  only two nonequivalent oxygen sites. The presence of apical oxygen ions in a fraction of interstitial sites that should be vacant is another source of lattice disorder \cite{Radaelli_1994}. Moreover, as-grown electron-doped cuprate samples are not superconducting and require an additional annealing step \cite{Tokura_1989}. Consequences of the annealing process and its effect on disorder remain under debate~\cite{Radaelli_1994,Schultz_1996,Riou_2004,Richard_2004,Richard_2004_2,Boulanger_2022}. More recently, the scattering of phonons on defects in the presence of short-range spin fluctuations was proposed as a mechanism for the large phonon thermal Hall effect observed in the electron-doped cuprates \cite{Boulanger_2022}. This context brings a renewed interest in the study of the interplay of disorder and spin fluctuations in these materials. 

From a theoretical standpoint, the interplay of disorder, spin fluctuations and hot spots has been studied phenomenologically in Ref.~\cite{Lin_Millis_2011}, generalizing  earlier phenomenological approaches used in the pure case in one and higher dimensions~\cite{Lee_Rice_Anderson_1973,sadovskii1979exact,Schmalian_Pines_Stojkovic_1999,Tchernyshyov_1999}. However this phenomenological approach involves many fit parameters and, in addition, no clear criterion was provided for the opening of a pseudogap. Disorder has been the subject of many other studies, as reviewed in Ref. \cite{Alloul_2009}. For instance, the effect of disorder in the Hubbard model has been studied with several methods~\cite{Janis_1993, Singh_1998, Byczuk_2009,Trivedi_2005,Chen_2009,Kontani_2006,Kontani_2008,Rosch:1999,Beal-Monod_1986} but none of them addressed the question of the effect of disorder on $T^*$, on the QCP, and on the criterion for the appearance of hot spots (AFM pseudogap). 

Here, we address this question in the $2D$ Hubbard model by generalizing the two-particle self-consistent approach (TPSC) using the impurity averaging technique. We work in the clean limit, neglecting weak localization effects~\cite{Altshuler_1980,Beal-Monod_1985}. Since hot spots appear as a precursors of an AFM ground state in $2D$, the Mermin-Wagner theorem needs to be satisfied, a requirement fulfilled by TPSC \cite{Vilk_1997}. We first show that disorder displaces $T^*$ and the AFM QCP towards a smaller filling and decreases AFM fluctuations. Then, we explain how an AFM pseudogap can be observed in the presence of disorder even when the Vilk criterion is not satisfied.

\section{Model and method}
\subsection{Model}
\label{sec:model}
We consider the one-band Hubbard model in $2D$
\begin{align}
    H = \sum_{\mathbf{k}}\sum_\sigma \epsilon_\mathbf{k}c^{\dagger}_{\mathbf{k},\sigma}c_{\mathbf{k},\sigma}+ U\sum_{i} n_{i\uparrow}n_{i\downarrow}, \label{eq:hubbard}
\end{align}
where $\epsilon_{\mathbf{k}} = -2t(\cos(k_x)+\cos(ky)) - 4t'\cos(k_x)\cos(k_y) - 2t''(\cos(2k_x)+\cos(2k_y))$ is the non-interacting dispersion, $c^{(\dagger)}_{\mathbf{k}\sigma}$ annihilates (creates) an electron of wave vector $\mathbf{k}$ and spin $\sigma$, $n_{i\sigma}$ is the number operator at site $i$ for spin-$\sigma$ electrons, and $U$ gives the strength of on-site interactions. 
To study a realistic case, we use first-, second- and third-neighbor hoppings $t=1$, $t'=-0.175$, $t''=0.05$ as well as $U=5.75$. These parameters yield agreement between theory and experiment for the ARPES spectrum of electron-doped cuprate NCCO for fillings larger than $n=1.1$ \cite{Kyung_2004}. We use units where $\hbar=1$ and $k_B=1$. 

\subsection{Method: the two-particle self-consistent approach with disorder}
\label{sec:method}
The two-particle self-consistent (TPSC) approach is a non-perturbative method for the Hubbard model that is valid from weak to intermediate interaction regimes ($U\leq 6t$) \cite{Vilk_1997}. This method respects not only the Mermin-Wagner theorem but also the Pauli principle. It is based on satisfying sum rules for the spin and charge susceptibilities. Working in Matsubara frequencies, these susceptibilities satisfy
\begin{align}
    \chi_{sp}(\mathbf{r}=\mathbf{0},\tau=0) &= \sum_{\mathbf{q},iq_n}\frac{\chi_0(\mathbf{q},iq_n)}{1-U_{sp}\chi_0(\mathbf{q},iq_n)/2} \nonumber \\
    &= n-2\langle n_{\uparrow}n_{\downarrow}\rangle, \label{eq:spinsum}\\
    \chi_{ch}(\mathbf{r}=\mathbf{0},\tau=0) &= \sum_{\mathbf{q},iq_n}\frac{\chi_0(\mathbf{q},iq_n)}{1+U_{ch}\chi_0(\mathbf{q},iq_n)/2} \nonumber \\
    &= n+2\langle n_{\uparrow}n_{\downarrow}\rangle - n^2,\label{eq:chargesum}\\
    \chi_0(\mathbf{q},iq_n) =-2\sum_{\mathbf{k},i\omega_n}&\mathcal{G}_0(\mathbf{k},i\omega_n)\mathcal{G}_0(\mathbf{k}+\mathbf{q},i\omega_n+iq_n),\label{eq:chi0}
\end{align}
where $\chi_0(\mathbf{q},iq_n)$ is the non-interacting density-density correlation function, $q_n=2n\pi T$ are bosonic Matsubara frequencies, $\omega_n=(2n+1)\pi T$ are fermionic Matsubara frequencies, $\mathcal{G}_0$ is the non-interacting Matsubara Green function, and $U_{sp}$ and $U_{ch}$ are spin and charge vertices respectively. Unlike in RPA, where $U_{sp}=U_{ch}=U$, the vertices that enter the spin and charge susceptibilities are renormalized in the TPSC approach. More specifically, the TPSC approach solves the local-moment sum rule \eref{eq:spinsum} self-consistently by finding the double occupancy $\langle n_\uparrow n_\downarrow \rangle$ that satisfies the ansatz
\begin{equation}
     U_{sp} = U \frac{\langle n_{\uparrow}n_{\downarrow}\rangle}{\langle n_{\uparrow}\rangle\langle n_{\downarrow}\rangle}.
     \label{eq:ansatz}
\end{equation}
The spin and charge susceptibilities are then used to calculate the self-energy $\Sigma$ that enters the interacting Green function $\mathcal{G}$
\begin{align}
    \Sigma(\mathbf{k},i\omega_n) &= \frac{U}{8}\frac{T}{N} \sum_{\mathbf{q},iq_n} \left [ 3U_{sp}\chi_{sp}(\mathbf{q},iq_n) \right. \nonumber \\
    &+ \left . U_{ch}\chi_{ch}(\mathbf{q},iq_n)\right]\mathcal{G}_0(\mathbf{k}+\mathbf{q},i\omega_n+iq_n), \label{eq:sigma2}
\end{align}
\begin{equation}
    \mathcal{G}(\mathbf{k},i\omega_n) = \frac{1}{i\omega_n - \epsilon_{\mathbf{k}}+\mu-\Sigma(\mathbf{k},i\omega_n)}, \label{eq:g2}
\end{equation}
where $\mu$ is the chemical potential that gives the correct filling computed with the interacting Green function.

To include the effect of disorder, we use the average over impurities technique \cite{Rickayzen_1980}. This results in adding a disorder self-energy to the non-interacting Green function 
\begin{align}
    &\mathcal{G}_0^{dis}(\mathbf{k},i\omega_n) = \frac{1}{i\omega_n-\epsilon_{\mathbf{k}}+\mu^{dis}_0-\Sigma^{dis}(i\omega_n)}, \label{eq:g0dis}\\
    &\Sigma^{dis}(i\omega_n) = -\frac{i}{2\pi\tau}\left (\arctan\left(\frac{\zeta_{max}}{\omega_n}\right)-\arctan\left(\frac{\zeta_{min}}{\omega_n}\right)\right),\label{eq:sigmadis}
\end{align}
where $\zeta_{min(max)}$ is the minimal (maximal) value of $\epsilon_{k}-\mu_0$, and $\mu^{dis}_0$ and $\mu_0$ are the chemical potentials that give the correct fillings computed with the non-interacting Green functions with and without disorder respectively. The parameter that sets the disorder level is the lifetime $\tau$, which is finite when disorder is present and goes to infinity in the case without disorder. To obtain the disorder self-energy of \eref{eq:sigmadis}, we neglect the vertex corrections in the diagram for impurity scattering and only consider contributions near the Fermi level \cite{Rickayzen_1980}. The approximation we use to obtain \eref{eq:g0dis} and \eref{eq:sigmadis} is only valid in the clean limit, namely when $k_F v_F \tau \gg 1$, where $v_F \tau$ is the mean free path $l$. In our model, this corresponds approximately to $\tau > 1$. In practice, we add disorder to our TPSC calculations by replacing the non-interacting Green functions in \eref{eq:chi0} with the disorder Green function $\mathcal{G}_0^{dis}$ defined in \eref{eq:g0dis}. We neglect vertex corrections due to disorder when calculating $\chi_0^{dis}$. We have verified that vertex corrections that are necessary to obtain the diffusion pole near $\mathbf{q}=0$ become less important near the AFM wave vector $\mathbf{Q}=(\pi,\pi)$ of interest to us~\cite{Lin_Millis_2011}. The disorder self-energy defined in \eref{eq:sigmadis} is also included in the interacting Green function. Vertex corrections due to disorder could also be present in the expression of the self-energy, but we also neglect them to be consistent with the calculation of susceptibilities. The TPSC ansatz \eref{eq:ansatz} remains unchanged, but the double occupancy should now be viewed as averaged over disorder.

\section{Results and discussion}
\subsection{Effect of disorder on $T^*$ and on the QCP} 
\label{sec:Tstar}
\begin{figure}
    \centering
    \includegraphics[width=0.85\columnwidth]{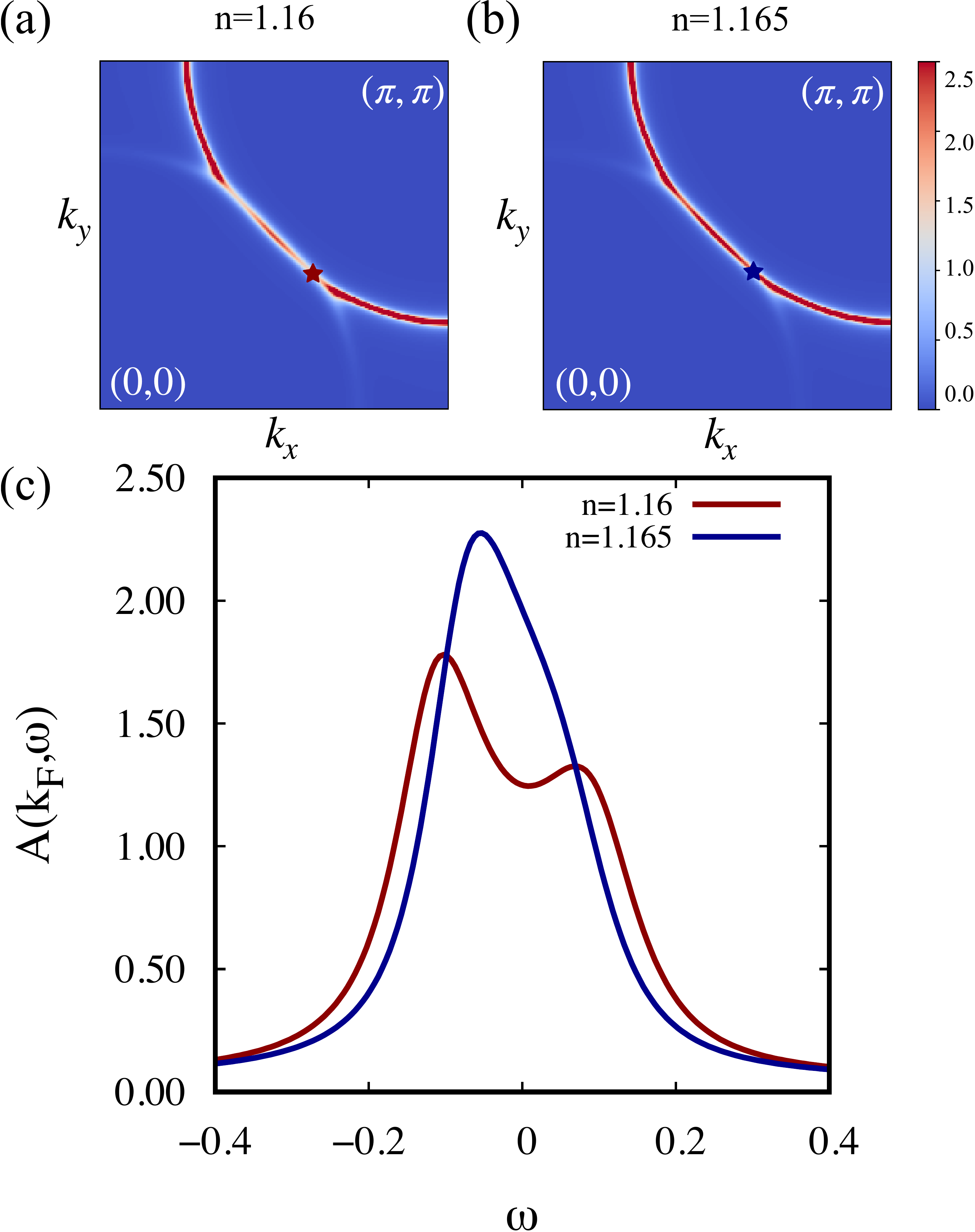} 
    \caption{Spectral weight computed with our NCCO-model parameters for $T=0.02$ and $\tau=25$ to illustrate how we define the AFM pseudogap temperature $T^*$. Figures (a) and (b) show $A(\mathbf{k},\omega=0)$, the spectral weight at $\omega=0$ as a function of wave vector $\mathbf{k}$ in a quarter of the Brillouin zone at fillings $n=1.16$ and $n=1.165$ respectively. The stars are placed at one of the two hot spots $\mathbf{k}_F$ where the Fermi surface crosses the AFM Brillouin zone. In (a), the AFM pseudogap is clearly visible at the hot spots for n=1.16. Figure (c) shows $A(\mathbf{k}_F,\omega)$, the spectral weight as a function of the frequency $\omega$ at the hot spots $\mathbf{k}_F$ defined in the panels (a) and (b). We see that, at this value of the temperature, a loss of spectral weight occurs at $\omega=0$ for $n=1.16$, but not for $n=1.165$, which defines the boundary of the AFM pseudogap at this temperature.}
    \label{fig:akw}
\end{figure}

We first investigate the effect of disorder on the phase diagram of our model. To do so, we do calculations at fixed temperatures for multiple values of the filling $n$ and obtain the spectral weight at the hot spot $A(\mathbf{k}_F,\omega)$ from Padé analytic continuation \footnote{For the Padé analytic continuation, we use the PyPade open-source implementation \href{https://www.physique.usherbrooke.ca/code_source/published_source_codes.html}{available online}, based on the algorithm presented in Ref. \cite{Vidberg_1977}.}. We define the boundary of the AFM pseudogap as the lowest filling at which we see a loss of spectral weight at the hot spot, as illustrated in \fref{fig:akw}. We restrict our calculations to fillings larger than $n=1.1$ due to our choice of band parameters and $U$. To include disorder, we use values of $\tau$ based on experimental measurements. A Dingle temperature $T_D = 14$ K was measured using Shubnikov-de Haas oscillations for NCCO at optimal doping ($n=1.15$) \cite{Helm_2013}. From $T_D = \hbar/2\pi k_B \tau$ and taking $t=350$ meV, we find $\tau=46$ in units of $1/t$. We also use $\tau=25$ as a way to consider samples that could be more disordered. For instance, direct resistivity measurements yield $l\simeq25a$ for thin films of PCCO at optimal doping \cite{Fournier_1998, Fournier_2015}. Taking $l=v_F\tau$, $a=4$\AA~\cite{Fournier_2015} and $v_F=2$ eV\AA~\cite{Horio_2020}, we find $\tau\simeq17.5$ in units of $1/t$, once again with $t=350$ meV.  In the range of temperatures and fillings we consider, the values $\tau=46$ and $\tau=25$ considered here correspond approximately to mean free paths $l\simeq120a$ and $l\simeq 65a$ respectively in our calculations. 

\begin{figure}
    \centering
    \includegraphics[width=0.5\columnwidth]{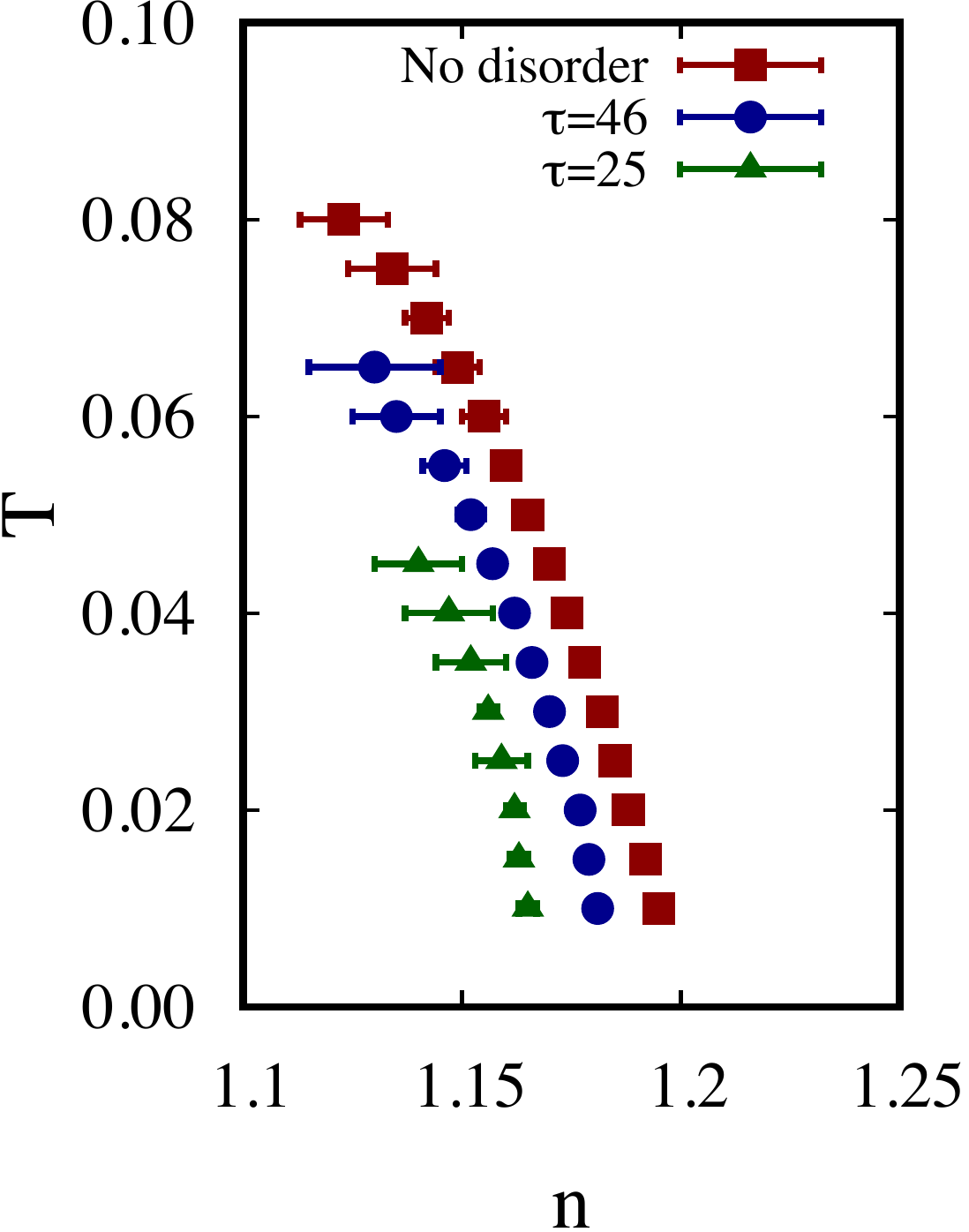} 
    \includegraphics[width=0.401\columnwidth]{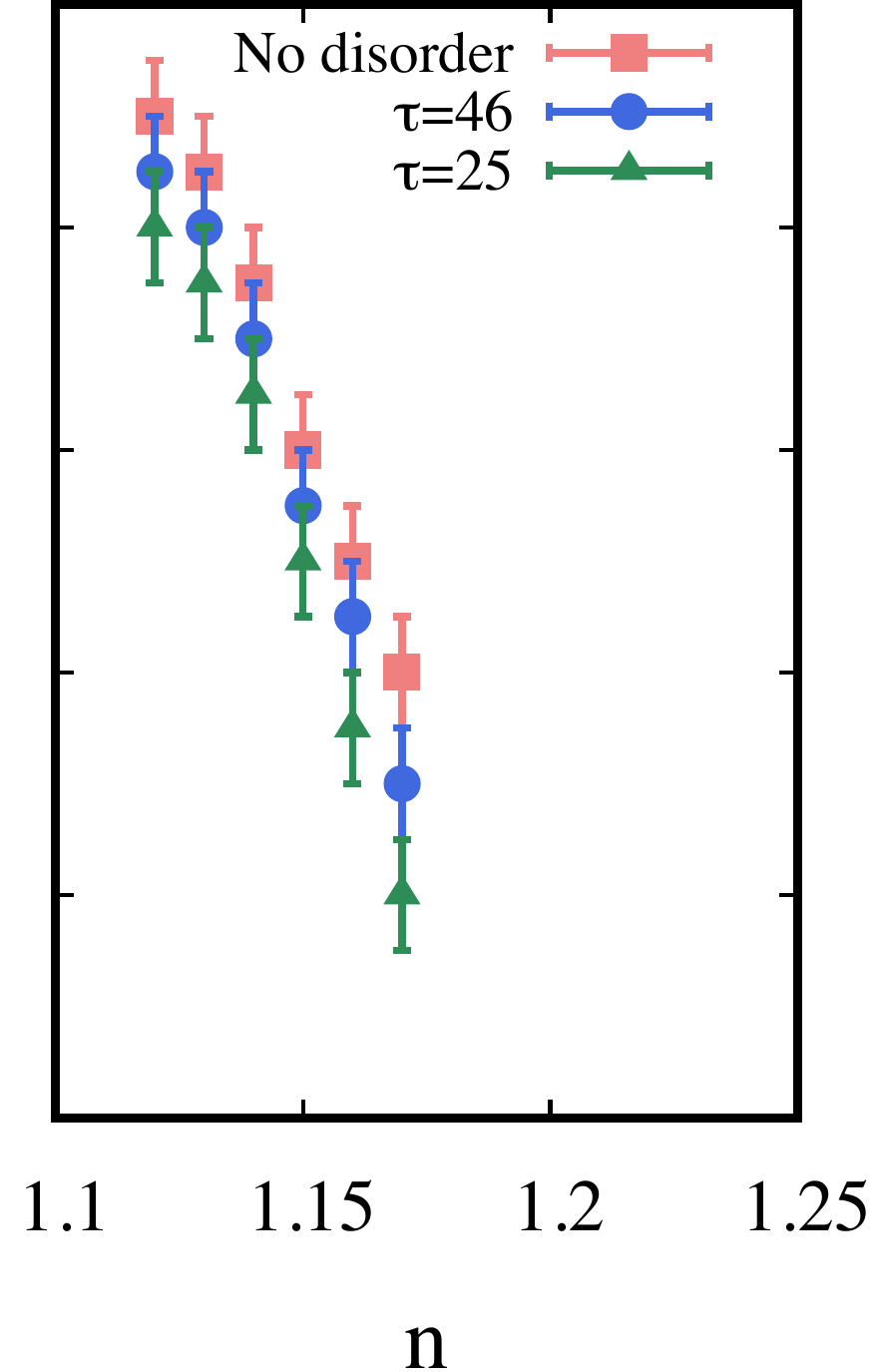} 
    \caption{Antiferromagnetic pseudogap temperature $T^*$ without disorder (red squares), with disorder $\tau=46$ (blue circles) and with disorder $\tau=25$ (green triangles). Left panel shows $T^*$ obtained from the spectral weight at the hot spots $A(\mathbf{k}_F,\omega)$. Right panel shows $T^*_{C_U}$ obtained from the maximum value of the specific heat over temperature $C_{U,n}/T$ at each filling.}
    \label{fig:phasediag}
\end{figure}

The resulting phase diagrams are shown in the left panel of \fref{fig:phasediag}. We compare the three cases mentioned above: no disorder, $\tau=46$ and $\tau=25$. Our calculations show that increasing disorder pushes the $T^*$ line towards lower fillings. As shown in \appref{sec:AFMQCP}, in our calculations for this model, an AFM QCP coincides with the filling where $T^*$ goes to $0$. We find that this AFM QCP is also displaced to a smaller value of the filling when disorder increases. This is expected because the width of the Fermi surface caused by disorder decreases nesting hence disfavouring AFM fluctuations.

Another way to construct the phase diagram is to study the specific heat. We obtain the electron-electron interaction contribution to the specific heat $C^{e-e}_{n,U}$ from differentiation of the free-energy obtained from coupling-constant integration   
\begin{equation}
    C^{e-e}_{n,U}(T,n,U) = - T \int_{0}^{U}dU'\left (\frac{\partial^2 \langle n_{\uparrow} n_{\downarrow} \rangle}{\partial T^2}\right )_{n, U'}.
    \label{eq:cU}
\end{equation}
In TPSC, coupling-constant integration is the most reliable way to obtain the free energy~\cite{Roy}. Details of this calculation are given in \appref{sec:specificheat}. 
In $2D$, the Mermin-Wagner theorem prevents long-range AFM order at finite temperature. However, previous calculations in the weak-coupling $2D$ Hubbard model have shown low-temperature peaks in the specific heat that could correspond to the energy scale of short-range spin fluctuations that can be associated with $T^*$ \cite{Roy}. We calculate $C^{e-e}_{n,U}$ as a function of the temperature $T$ for multiple values of $n$. We define $T_{C_U}^*$ as the temperature at which $C^{e-e}_{n,U}/T$ is maximal for a given filling $n$. The results for $T_{C_U}^*$ are shown in the right panel of \fref{fig:phasediag} for the cases without disorder, and for the two disordered cases $\tau=46$ and $\tau=25$. We find that there is a good agreement between the trend in $T^*$ defined previously from the spectral weight and $T_{C_U}^*$, as shown in \fref{fig:phasediag}, though the effect of disorder seems more prominent in the spectral weight calculations than in the specific heat, as expected from the fact that specific heat involves all degrees of freedom, not just hot spots. 

To further understand the effect of disorder on the AFM fluctuations, we now turn to the spin correlation length. In the regime where spin fluctuations are strong, and the characteristic spin-fluctuation frequency is smaller than temperature (the renormalized classical regime), the spin susceptibility can be written in the Ornstein-Zernicke form \cite{Vilk_1997}
\begin{equation}
    \chi_{sp}(\mathbf{q},0) \simeq \frac{1}{U_{sp}\xi_0^2}\frac{2}{(\mathbf{q}-\mathbf{Q})^2 +\xi_{sp}^{-2}},
    \label{eq:chisp}
\end{equation}
where $\xi_{sp}$ is the spin correlation length, while the bare particle-hole correlation length~$\xi_0$ is defined by 
\begin{equation}
    \xi_{0}^2 = -\frac{1}{2\chi_0(\mathbf{Q},0)}\left . \frac{\partial^2 \chi_0(\mathbf{q})}{ \partial q_x^2}\right|_{\mathbf{q}=\mathbf{Q}}.
    \label{eq:xi02}
\end{equation}
 In \eref{eq:chisp}, the spin correlation length is defined as
\begin{equation}
    \xi_{sp} = \xi_0\sqrt{\frac{U_{sp}}{\delta U}},
    \label{eq:xisp}
\end{equation}
where
\begin{align}
    \delta U &= U_{mf}-U_{sp},\nonumber\\
    U_{mf} &= \frac{2}{\chi_0(\mathbf{Q},0)}.
    \label{eq:deltaU}
\end{align}
Here, $U_{mf}$ is the critical value of $U_{sp}$ that would lead, within RPA, to an AFM phase transition that is forbidden in $2D$ at finite temperature.

In practice, we perform the TPSC calculation to find $U_{sp}$ and $\chi_{sp}(\mathbf{q},i\omega_n)$ self-consistently. We then calculate the spin correlation length $\xi_{sp}$ from the width at half maximum of $\chi_{sp}(\mathbf{q},0)$ around $\mathbf{Q}=(\pi,\pi)$. We also obtain $U_{mf}$ from the TPSC calculation. We use all these results to extract $\xi_0$ using \eref{eq:xisp}.

\begin{figure}
    \centering
    \includegraphics[width=0.85\columnwidth]{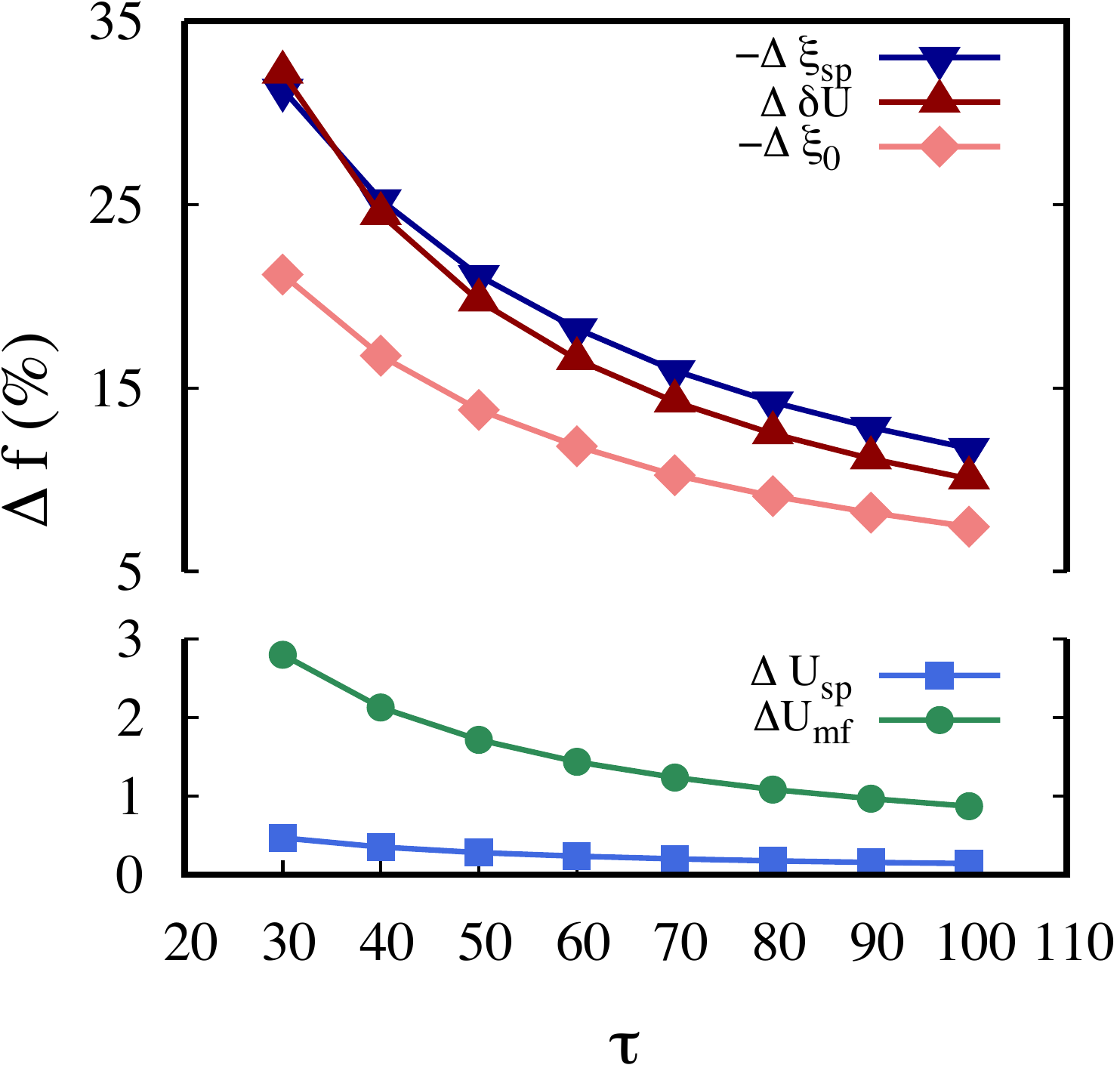} 
    \caption{Effect of disorder on the spin fluctuations. Calculations are done at $T=0.04$ and $n=1.20$, the filling associated with the AFM QCP for the case without disorder. As $\tau$ increases, the disorder level decreases. Here, we use the notation $\Delta f = (f^{\mathrm{dis}}-f)/f$ to signify the relative deviation in percentage between the results with and without disorder. Blue squares and green circles show respectively the relative deviation of $U_{sp}$ and $U_{mf}$ due to disorder. Both quantities are only slightly increased due to disorder. Red diamonds, dark-blue triangles and dark-red triangles show respectively the relative deviation of $\xi_0$, $\xi_{sp}$ and $\delta U$ due to disorder. The large increase in $\delta U$ and decrease in $\xi_0$ due to disorder suppress the spin correlation length \eref{eq:xisp}.}
    \label{fig:vartau}
\end{figure}

We show the effect of disorder on $\xi_{sp}$, $\delta U$, $U_{sp}$, $U_{mf}$ and $\xi_0$ in \fref{fig:vartau}, where we decrease the disorder from $\tau=30$ to $\tau=100$. All the calculations in this plot are done for $T=0.04$ at $n=1.20$, the filling associated with the AFM QCP of the model without disorder. We find that the spin correlation length $\xi_{sp}$ is suppressed when disorder increases. This comes from the behavior of both $\delta U$ and $\xi_0$ in the expression for the spin correlation length \eref{eq:xisp}. Indeed, there is a significant increase of $\delta U$ since the QCP moves to smaller doping than $n=1.2$ with disorder, concommitant with a decrease of $\xi_0$ from the rounding of $\chi_0$ by disorder. In contrast, $U_{sp}$ and $U_{mf}$ are only slightly affected by disorder. In \appref{sec:chi0pheno}, we show that these effects on $U_{mf}$ and $\xi_0$ can be understood qualitatively using a phenomenological approach. 

We note that the total scattering rate cannot be computed as the sum of the elastic part from disorder $\Gamma_{el.}$ and an inelastic part obtained from a pure, non-disordered calculation: We have the inequality $\Gamma_{tot} \neq \Gamma_{el.}+\Gamma^{\mathrm{pure}}_{inel.}$, since $\Gamma_{inel.}$ is modified by disorder even in the clean limit we consider here. Indeed, as seen from \eref{eq:sigma2}, the TPSC self-energy, and hence the inelastic scattering rate, directly depends on the spin correlations through $U_{sp}$ and $\chi_{sp}$ that are modified by disorder. This breakdown of Matthiessen's rule was also noted in Ref.~\cite{Rosch:1999}.

\subsection{Criterion for the antiferromagnetic pseudogap} 
\label{sec:criterion}
\begin{figure}
    \centering
    \includegraphics[width=0.65\columnwidth]{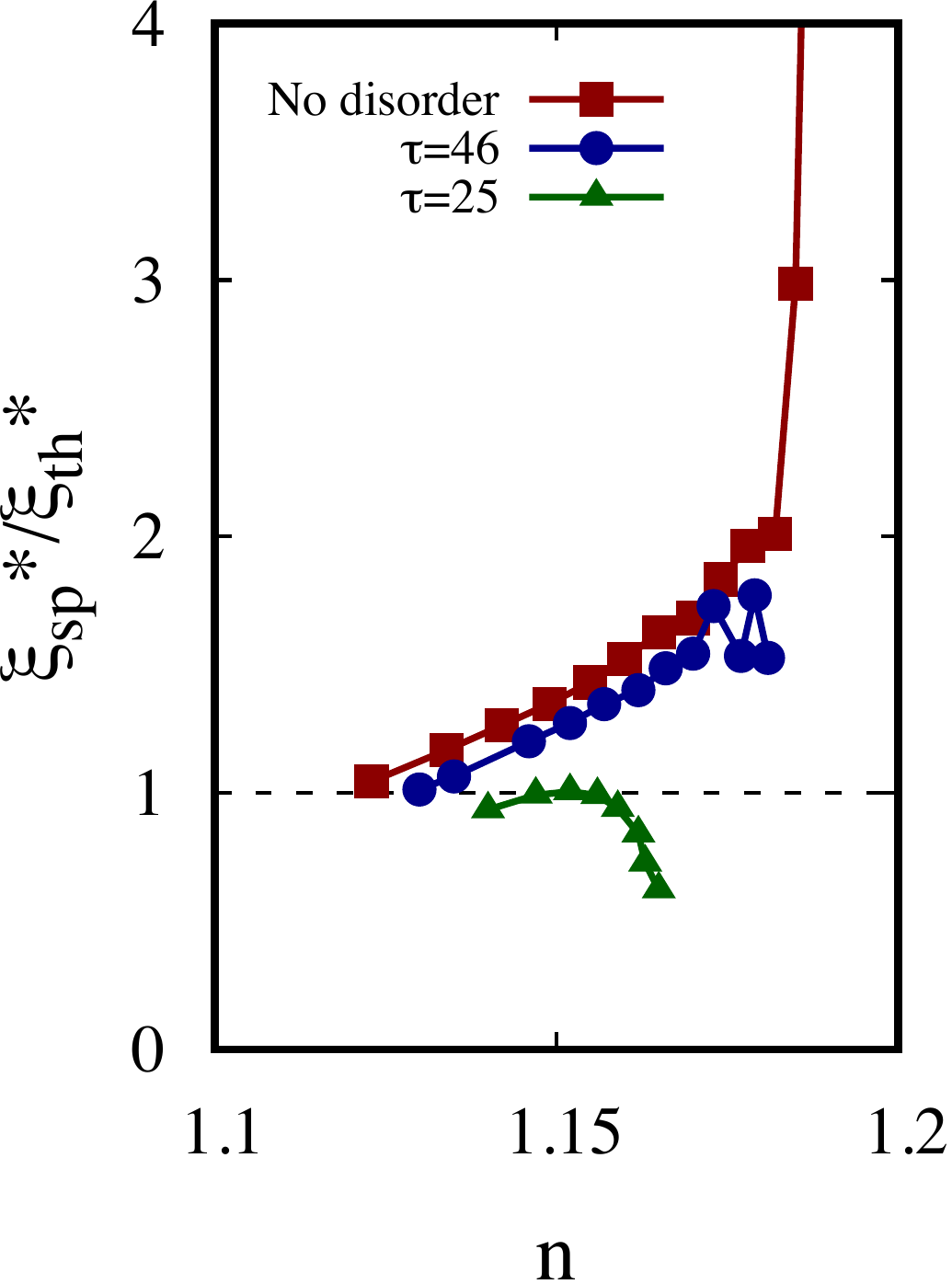} 
    \caption{Ratio of the spin correlation length $\xi_{sp}$ and the thermal de Broglie wavelength $\xi_{th}$ evaluated at $T^*$. The Vilk criterion $\xi_{sp}^*\geq\xi_{th}^*$, shown as the dashed line in the plot, is satisfied in the cases without disorder (dark-red squares) and with $\tau=46$ (dark-blue circles). For a higher disorder level of $\tau=25$ (green triangles), the criterion is not satisfied at fillings near the AFM QCP, even though an AFM pseudogap can be seen in the spectral weight (\fref{fig:akw}) in this range of parameters.
    }
    \label{fig:criterexisp}
\end{figure}

As mentioned previously, multiple methods for the 2D Hubbard model have shown that the AFM pseudogap for weak interactions opens at the hot spots when the Vilk criterion $\xi_{sp}>\xi_{th}$ is realized \cite{Schafer_2021}. Here, we investigate whether this criterion still holds in the presence of disorder. We show in \fref{fig:criterexisp} (a) the ratio $\xi_{sp}/\xi_{th}$ evaluated at the AFM pseudogap temperatures $T^*$ shown in the left panel of \fref{fig:phasediag}. In the case without disorder, the criterion is satisfied along $T^*$, as expected from previous TPSC calculations. The same is true for the $\tau=46$ case. In both of these calculations, we find that, as the filling increases towards the AFM QCP, the ratio $\xi_{sp}/\xi_{th}$ rapidly becomes larger than $1$. This must be because, contrary to the underdoped case where $\xi_{sp}$ increases much more rapidly with $T$ than $\xi_{th}$, at the AFM QCP both $\xi_{sp}$ and $\xi_{th}$ have the same $1/T$ temperature dependence \cite{Bergeron_2012}.  The $1/T$ dependence of $\xi_{sp}$ at the QCP comes from pseudonesting as seen in Fig.~3 of Ref.~\cite{Bergeron_2012}. 

In stark contrast, \fref{fig:criterexisp} shows that, in the $\tau=25$ case, the ratio $\xi_{sp}/\xi_{th}$ falls below $1$ at $T^*$ as the filling increases towards the AFM QCP. 
Hence the Vilk criterion does not have to be respected for an AFM pseudogap to open at the hot spots when sufficient disorder is added to the model.

To understand the difference between the $\tau=25$ case and the other two cases, we follow Ref. \cite{Vilk_1997} to obtain an analytic expression for the self-energy $\Sigma_{cl}(\mathbf{k}_F,0)$ at the hot spots from a phenomenological approach. In the renormalized classical regime, the dominant contribution to the self-energy  \eref{eq:sigma2} comes from spin fluctuations at $q_n=0$. Furthermore, the spin susceptibility can be written in this regime using the asymptotic form \eref{eq:chisp}.  Details of the calculation can be found in \appref{sec:selfcf}. In the end, we obtain the imaginary part of the self-energy 
\begin{equation}
    \mathrm{Im}\Sigma_{cl}(\mathbf{k}_F,0,\tau) =   -\frac{3U}{8\pi^2\xi_0^2(\tau)}\frac{\xi_{sp}}{\xi_{th}} F\left(\frac{\xi_{sp}}{v_F\tau}\right),
    \label{eq:imSelfPheno}
\end{equation}
where
\begin{equation}
    F(x) =  \left \{ \begin{matrix}\frac{\pi}{2} & & x=0~\textrm{(No~disorder)},\\
        \frac{\mathrm{arccosh}(x)}{\sqrt{x^2-1}} &  & x > 0.
    \end{matrix} \right.
    \label{eq:imSelfPhenoFDis}
\end{equation}
This result agrees with Ref.~\cite{Lin_Millis_2011}. The effect of disorder on the self-energy is not obvious since disorder affects not only the function $F(x)$ defined in \eref{eq:imSelfPhenoFDis}, but also the prefactor $\xi_0^{-2}$. As shown in \fref{fig:vartau}, $\xi_0$ is suppressed by disorder. In \appref{sec:chi0pheno}, we show analytically that $\xi_0$ decreases with disorder in the clean limit whenever the non-disordered $\xi_0$ is small compared with the thermal de Broglie wavelength, which is always satisfied in the cases we study here. Numerically, $\xi_0^2$ varies in the range $0.65<\xi_{0}^2(\tau)/\xi_0^2(\tau\rightarrow \infty)<0.9$ when we vary $\tau$ between $30$ and $100$ at $n=1.20$ and $T=0.04$. 

To understand how the above result for $\mathrm{Im}\Sigma_{cl}(\mathbf{k}_F,0,\tau)$ modifies the Vilk criterion, let us first note that the AFM pseudogap at the hot spot $\mathbf{k}=\mathbf{k}_F$ opens when $\mathrm{Im}\Sigma_{cl}(\mathbf{k}_F,0,\tau)$ becomes large. In the case without disorder this happens when $\xi_{sp}$ becomes larger than the thermal de Broglie wavelength $\xi_{th}$, as seen from \eref{eq:imSelfPheno} and \eref{eq:imSelfPhenoFDis} at $x=0$. This is the original formulation of the Vilk criterion. Let us then take
\begin{equation}
    |\mathrm{Im}\Sigma_{cl}(\mathbf{k}_F,0,\tau)|\geq\frac{3U}{16\pi\xi_0^2(\tau\rightarrow \infty)}.
    \label{eq:critereSigma}
\end{equation}
as a criterion on the minimal value of the imaginary part of the self-energy for the AFM pseudogap to open. We note that, even if we deduced \eref{eq:critereSigma} from the case without disorder $(\tau\rightarrow \infty)$, it also holds for cases with finite values of the lifetime $\tau$.

Then, from \eref{eq:critereSigma} and \eref{eq:imSelfPheno}, we find a new formulation of the Vilk criterion that takes into account the effects of disorder. An AFM pseudogap opens at the hot spots when
\begin{equation}
    \frac{\xi_{sp}}{\xi_{th}} \geq \frac{\pi}{2}\frac{\xi_0^2(\tau)}{\xi_0^2(\tau\rightarrow \infty )}\frac{1}{F(\frac{\xi_{sp}}{v_F\tau})},
    \label{eq:critereRatio}
\end{equation}
from which the original Vilk criterion criterion is recovered by taking $\tau\rightarrow \infty$.

\begin{figure}
    \centering
    \includegraphics[width=0.9\columnwidth]{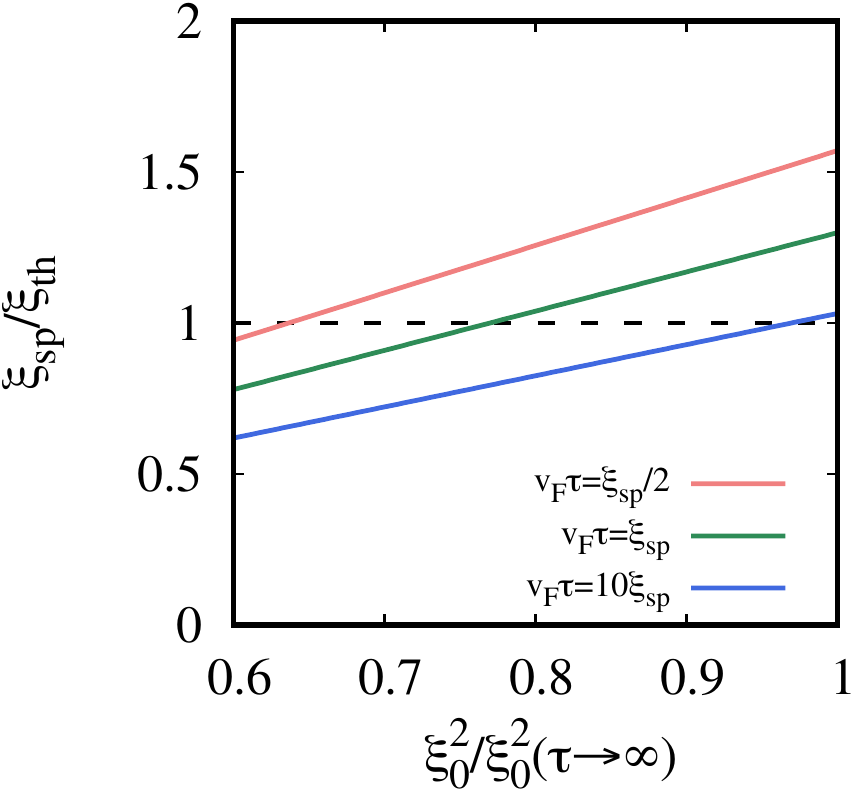} 
    \caption{Minimal value of the ratio $\xi_{sp}/\xi_{th}$ at wich an AFM pseudogap is expected to open at the hot spots from \eref{eq:critereRatio}. The ratio $\xi_0^2/\xi_0^2(\tau\rightarrow \infty)$ is taken as a free parameter in the analytic calculation. The black dashed line corresponds to the original Vilk criterion. The three other lines, calculated with disorder: $v_F\tau = \xi_{sp}/2$ (pink line), $v_F\tau = \xi_{sp}$ (green line), $v_F\tau = 10\xi_{sp}$ (blue line), show the value of $\xi_{sp}/\xi_{th}$ at which an AFM pseudogap opens depending on the value of $\xi_0^2/\xi_0^2(\tau\rightarrow \infty)$.}
    \label{fig:critereRatio}
\end{figure}

We plot the ratio obtained from the strict equality in \eref{eq:critereRatio} in \fref{fig:critereRatio} as a function of $\xi_0^2(\tau)/\xi_0^2(\tau\rightarrow \infty)$. The Vilk criterion corresponds to the constant, black dashed line at $\xi_{sp}/\xi_{th}=1$. We also consider three cases with varying levels of disorder through the ratio $\xi_{sp}/v_F\tau$ that enters the function $F$ defined in \eref{eq:imSelfPhenoFDis}: $v_F\tau=\xi_{sp}/2$ (pink line), $v_F\tau=\xi_{sp}$ (green line) and $v_F\tau=10\xi_{sp}$ (blue line). 

From our analytic calculations and their illustration in \fref{fig:critereRatio}, we find that the effect of disorder on the AFM pseudogap is subtle. The first modification of the self-energy by disorder depends on the ratio of $\xi_{sp}$ to the mean-free path $v_F\tau$, that appears in the argument of $F(x)$ defined in \eref{eq:imSelfPhenoFDis}. This seems to favour the opening of the AFM pseudogap when $\xi_{sp}\geq \xi_{th}$, similarly to the no-disorder case. However, disorder also suppresses $\xi_0^2$, which has the reverse effect on the criterion for the AFM pseudogap. Taken together, these disorder effects mean that an AFM pseudogap can exist even when $\xi_{sp}<\xi_{th}$, as shown in \fref{fig:critereRatio}.

\section{Conclusion} 
We showed that when disorder is included, an AFM pseudogap can open at the hot spots in the $2D$ Hubbard model even when $\xi_{sp}$ is less than $\xi_{th}$, contrary to the Vilk criterion. This may be an explanation for the experimental observation of a pseudogap even with a small $\xi_{sp}$ near optimal doping~\cite{Motoyama_2007}. Hence, systematic experimental investigation of the effect of disorder on hot spots is called for. A full comparison to experiment may however require including the effect of the interplay between superconducting and AFM fluctuations~\cite{Bourbonnais_Sedeki_2009}. An additional step would include a wave vector dependent elastic scattering rate, which could be more appropriate for the study of out-of-plane disorder.

\section*{Acknowledgments} 
We are grateful to Patrick Fournier, Michel Gingras and Marie-Eve Boulanger for useful discussions. This work has been supported by the Natural Sciences and Engineering Research Council of Canada (NSERC) under grant RGPIN-2019-05312, by a Vanier Scholarship (C. G.-N.) from NSERC, by a USRA scholarship from NSERC (P.-A. G.), and by the Canada First Research Excellence Fund. Simulations were performed on computers provided by the Canadian Foundation for Innovation, the Minist\`ere de l'\'Education des Loisirs et du Sport (Qu\'ebec), Calcul Qu\'ebec, and Compute Canada.

\newpage


%

\appendix

\section{Calculation of the AFM QCP}
\label{sec:AFMQCP}
\begin{figure*}
     \centering
     \includegraphics[width=0.36\textwidth]{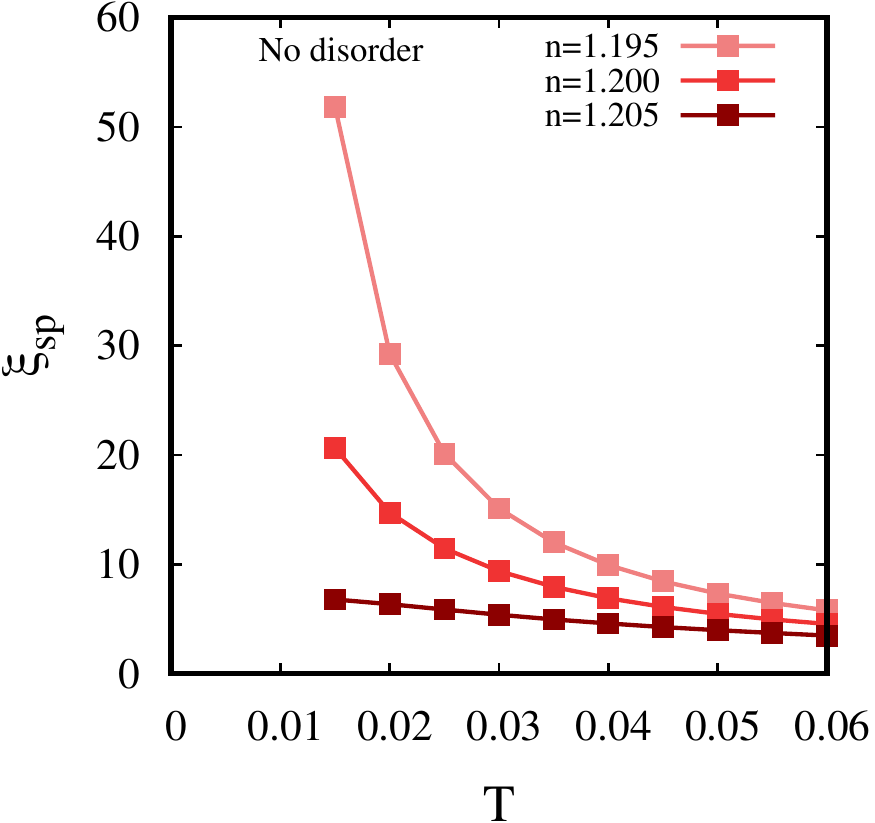}
     \includegraphics[width=0.3\textwidth]{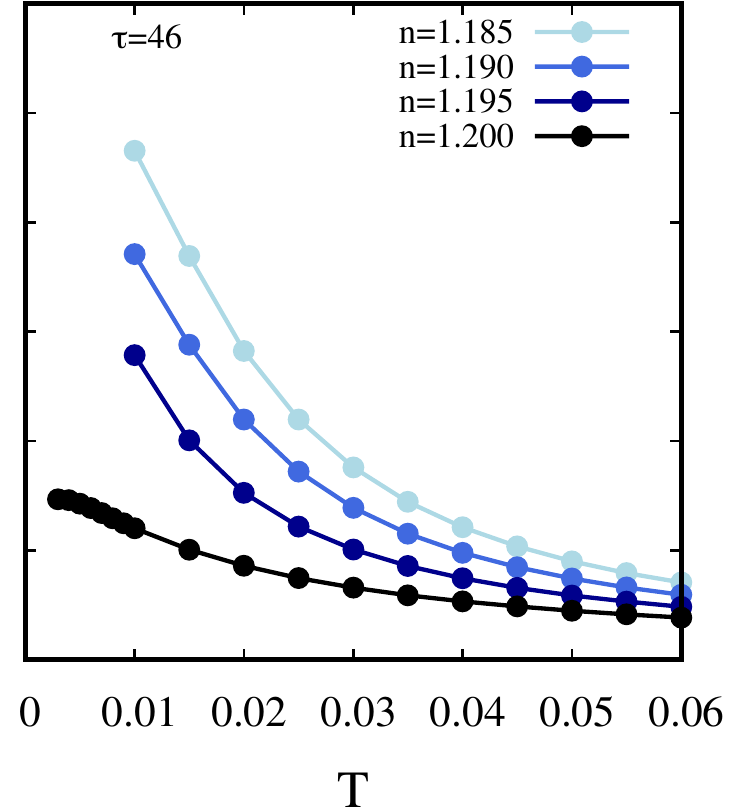}
     \includegraphics[width=0.3\textwidth]{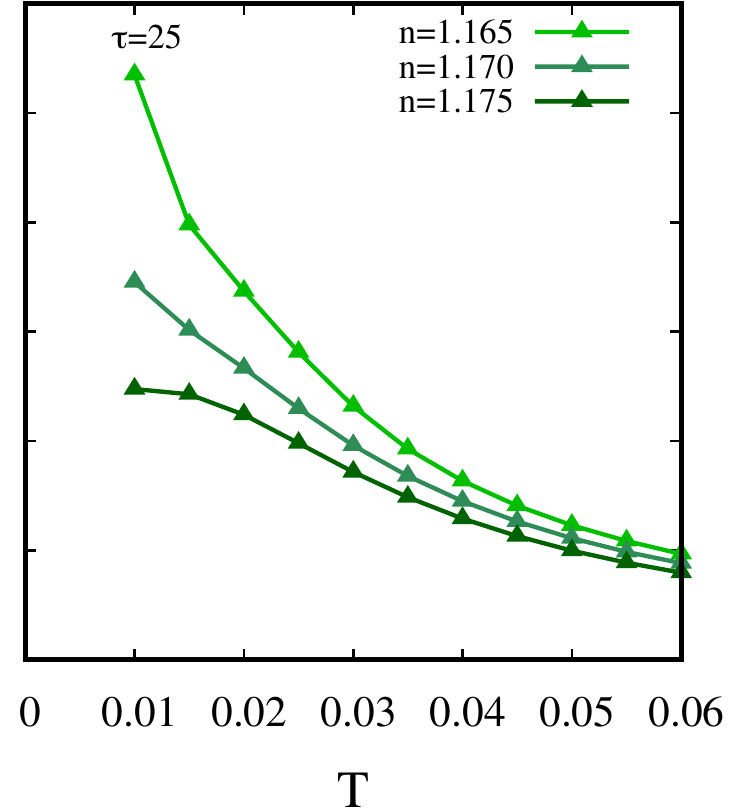}
     \caption{Temperature dependence of the spin correlation length as a function of the disorder level. Left panel shows the non-disordered case for fillings $n=1.195$, $n=1.200$ and $n=1.205$. The AFM QCP for this case is located around $n=1.200$, as seen from the behavior of the spin correlation length, which grows when $T\rightarrow0$ for this filling, but remains very small at low temperature for a slightly larger filling $n=1.205$. The middle panel shows the results for $\tau=46$ and the right panel for $\tau=25$, both cases with disorder. Similarly to the case without disorder, we find the AFM QCP for these systems by looking at the behavior of the spin correlation length at low temperature. We find that the QCP is around $n=1.195$ for $\tau=46$ and $n=1.170$ for $\tau=25$. All the results shown here are obtained from calculations converged with system size $1024\times 1024$ and for $2048$ Matsubara frequencies.}
     \label{fig:QCP_xisp}
 \end{figure*}

In our model, an AFM QCP is present at $T=0$ above half-filling. However, the TPSC approach we use to solve the problem works at finite temperature. We can still use our approach to determine the AFM QCP from the calculation of the spin correlation length at finite temperature. We calculate the spin correlation length as a function of temperature and filling. For fillings below the AFM QCP, the spin correlation length increases as temperature goes to zero. In contrast, for fillings above the AFM QCP, the spin correlation length should remain small or saturate as the temperature decreases. We show our results in \fref{fig:QCP_xisp} for the cases without disorder and with disorder ($\tau=46$ and $\tau=25$). We find that the AFM QCP is displaced towards a smaller filling when disorder increases, going from $n\simeq1.20$ without disorder to $n\simeq1.17$ for $\tau=25$. We also find that the AFM QCP obtained from the spin correlation calculations is consistent with the extrapolation to $T=0$ from the spectral weight estimate of the $T^*$ lines shown in \fref{fig:phasediag}.
    
\section{Specific heat calculation}
\label{sec:specificheat}
In this appendix, we show how we calculate the specific heat in our models with and without disorder. The results of these calculations were used to produce Fig. 2 in the main text.

We follow the coupling constant integration method, and more explicitly Ref. \cite{Roy} to compute the specific heat, starting from the definition of the grand canonical potential normalized per site:
\begin{equation}
    \Omega(T,\mu,U) = -\frac{T}{N_s} \ln\mathrm{Tr} \left [e^{-\beta \left(H - \mu N\right)}  \right ],
    \label{eq:grandpot}
\end{equation}
with the Hamiltonian $H$ of \eref{eq:hubbard}, $\beta$ the inverse temperature, and $N_s$ the number of sites. This potential can be calculated using an integral over $U$. Indeed, the derivative of the grand canonical potential with respect to $U$ simply yields the double occupancy
\begin{equation}
    \left ( \frac{\partial \Omega(T,\mu,U)}{\partial U} \right )_{T, \mu}= \langle n_{\uparrow} n_{\downarrow} \rangle.
\end{equation}
Then, integrating over the Hubbard $U$ gives
\begin{align}
    \int_{0}^{U}dU'&\left ( \frac{\partial \Omega(T,\mu,U)}{\partial U'} \right )_{T, \mu} =  \Omega(T,\mu,U) -  \Omega^0(T,\mu),\nonumber \\
    &\Rightarrow \Omega(T,\mu,U) = \Omega^0(T,\mu) + \int_{0}^{U}dU'\langle n_{\uparrow} n_{\downarrow} \rangle,
    \label{eq:intgrandpot}
\end{align}
where $\Omega^0(T,\mu)$ is the grand canonical potential calculated from the non-interacting Hamiltonian, and the double occupancy $\langle n_{\uparrow} n_{\downarrow} \rangle$ is a function of $U$ and $T$.
We then calculate the free energy normalized per site using a Legendre transform
\begin{equation}
    F(T,n,U) = \Omega(T,\mu(T,n,U),U) + \mu(T,n,U)n.
    \label{eq:legendreF}
\end{equation}
From the free energy, we obtain the following expression for the specific heat 
\begin{align}
    C_{n,U}(T,n,U) &= -T\left ( \frac{\partial^2 F(T,n,U)}{\partial T^2} \right )_{U, n},\nonumber \\
    = C^0_n(T,n) &- T \int_{0}^{U}dU'\left (\frac{\partial^2 \langle n_{\uparrow} n_{\downarrow} \rangle}{\partial T^2}\right )_{n, U'},
    \label{eq:chaleurspec}
\end{align}
where the non-interacting specific heat can be calculated from the energy per site:
\begin{equation}
C^0_n(T,n) = \frac{1}{N_s}\left (\frac{\partial E^0(T,n)}{\partial T}\right )_{n}.
\end{equation}
The electron-electron interaction contribution to the specific heat $C^{e-e}_{n,U}(T,n,U)$ is, as defined in the main text in \eref{eq:cU},
\begin{equation}
    C^{e-e}_{n,U}(T,n,U) = - T \int_{0}^{U}dU'\left (\frac{\partial^2 \langle n_{\uparrow} n_{\downarrow} \rangle}{\partial T^2}\right )_{n, U'}.
    \end{equation}
In practice, we compute the double occupancy $\langle n_{\uparrow} n_{\downarrow} \rangle$ at a fixed filling $n$ for multiple values of $U$ to perform a numerical integration using Simpson's rule. The derivatives with respect to temperatures are calculated using finite differences.

\section{Phenomenological approach for the Lindhard function}
\label{sec:chi0pheno}

The suppression of AFM fluctuations through disorder can be further understood using a phenomenological approach. The calculations are analogous to the Abrikosov-Gor'kov theory of non-magnetic impurities in superconductors. We start by focusing on the non-interacting density-density correlation function $\chi_0(\mathbf{Q},i\omega_0=0)$, where $\mathbf{Q}=(\pi,\pi)$ is the AFM wave vector, with and without disorder. We consider the specific case with only nearest-neighbor hopping ($t'=t''=0$) at half-filling ($\mu=0$). We study this case since we can then calculate $\chi_0(\mathbf{Q},i\omega_0=0)$ analytically without disorder
\begin{align}
    \chi_0(\mathbf{Q},i\omega_0=0) &= -2 \frac{T}{N}\sum_{\mathbf{k}, ik_n} \mathcal{G}_0(\mathbf{k},ik_n) \mathcal{G}_0(\mathbf{k}+\mathbf{Q},ik_n),\nonumber \\
    &= -2 \frac{T}{N}\sum_{\mathbf{k}, ik_n} \frac{1}{ik_n-\epsilon_{\mathbf{k}}}\frac{1}{ik_n-\epsilon_{\mathbf{k}+\mathbf{Q}}}.
\end{align}
In the square lattice with $t'=t''=0$, we find $\epsilon_{\mathbf{k}} = -\epsilon_{\mathbf{k}+\mathbf{Q}}$. Using this, we can first perform the sum over Matsubara frequencies, and then sum over the momentum $\mathbf{k}$. This yields
\begin{equation}
    \chi_0(\mathbf{Q},i\omega_0=0) = 2 \rho(\epsilon_F)\left [\ln{\left(\frac{2\Omega}{\pi T}\right)} +\gamma \right ],
    \label{eq:chi0PA}
\end{equation}
where $\Omega$ is an energy cutoff, $\rho(\epsilon_F)$ is the density of states at the Fermi level, and $\gamma$ is Euler's constant. Here we assumed that the density of states at the Fermi level is a constant for simplicity. In the presence of disorder the van Hove singularity is broadened.

With disorder, we instead have to calculate 
\begin{align}
    \chi_0^{\mathrm{dis}}(\mathbf{Q},i\omega_0=0) &= -2 \frac{T}{N}\sum_{\mathbf{k}, ik_n} \mathcal{G}_0^{\mathrm{dis}}(\mathbf{k},ik_n) \mathcal{G}_0^{\mathrm{dis}}(\mathbf{k}+\mathbf{Q},ik_n), \nonumber \\
    = -2 \frac{T}{N}\sum_{\mathbf{k}, ik_n}&\left [ \frac{1}{ik_n-\epsilon_{\mathbf{k}}-i\mathrm{sgn}(k_n)/(2\tau)} \right. \nonumber \\
    &\times  \left. \frac{1}{ik_n-\epsilon_{\mathbf{k}+\mathbf{Q}}-i\mathrm{sgn}(k_n)/(2\tau)} \right ].
\end{align}
Because of the self-energy due to disorder $i\mathrm{sgn}(k_n)/2\tau$, the sum over Matsubara frequencies cannot be performed analytically before the sum over wave vectors. To address this issue, we calculate
\begin{equation}
    \chi^{\mathrm{dis}}_0 =  \bar{\chi}^{\mathrm{dis}}_0 -  \bar{\chi}_0 + \chi_0,
\end{equation}
where $\bar{\chi}$ is calculated by performing the sum over wave vectors first, and $\chi$ is calculated by performing the sum over Matsubara frequencies first. We then find, with $\Omega$ the high-frequency cutoff,
\begin{align}
    \chi^{\mathrm{dis}}_0(\mathbf{Q},i\omega_0=0) &=  \chi_0(\mathbf{Q},0) -2\rho(\epsilon_F)a(0,4\pi T\tau)
    \label{eq:chi0disPA1}
\end{align}
with
\begin{equation}
    a(n,x) = \psi^{(n)}\left (\frac{1}{2}+\frac{1}{x}\right ) - \psi^{(n)}\left (\frac{1}{2}\right ),
    \label{eq:anx}
\end{equation}
where $\psi^{(n)}$ is the polygamma function of order $n$.  

In the clean limit, \eref{eq:chi0disPA1} can be approximated by
\begin{align}
    \chi^{\mathrm{dis}}_0(\mathbf{Q},i\omega_0=0) \simeq \chi_0(\mathbf{Q},i\omega_0=0) - \frac{\rho(\epsilon_F)\pi}{4T\tau}.
    \label{eq:chi0disPA}
\end{align}
This last result implies that AFM fluctuations are reduced by the addition of disorder. This analytic result is consistent with our previous numerical study of $U_{mf} = 2/\chi_0(\mathbf{Q},0)$, shown in \fref{fig:vartau}, which we found to increase with the level of disorder.

At the RPA level, the mean-field AFM transition temperature $T^{\mathrm{RPA}}_C$ is obtained from the Stoner criterion $1 = U\chi_0(\mathbf{Q},i\omega_0=0)/2$. Using this criterion, and taking $T_C$ and $T'_C$ as the RPA transition temperature with and without disorder respectively, we find
\begin{equation}
    T'_C \exp{\left[\frac{\pi}{8T'_C\tau}\right]} = T_C.
\end{equation}
Since we are interested in the clean limit, we can expand the exponential around $\tau\rightarrow \infty$, which gives
\begin{equation}
    T'_C  = T_C - \frac{\pi}{8\tau}.
    \label{eq:TcRPA}
\end{equation}
From \eref{eq:TcRPA}, we can see that disorder in the clean limit lowers the RPA mean-field critical temperature. 

To compute $\xi_0^2$, which is defined as 
\begin{equation}
    \xi_{0}^2 = -\frac{1}{2\chi_0(\mathbf{Q},0)}\left . \frac{\partial^2 \chi_0(\mathbf{q})}{ \partial q_x^2}\right|_{\mathbf{q}=\mathbf{Q}},
\end{equation}
we can do a similar calculation of $\chi_0(\mathbf{Q}+\mathbf{q})$ at small $\mathbf{q}$ and then obtain its second derivative. This yields
\begin{align}
    \frac{\xi_{0}^{dis~2}}{\xi_{0}^{2}} = &\frac{1}{\chi^{dis}_0(\mathbf{Q},0)}\times \nonumber \\
    &\left [ \chi_0(\mathbf{Q},0) -\frac{\rho(\epsilon_F)}{48}\frac{\xi_{th}^2}{\xi_0^2}a(2,4\pi T\tau)\right],
    \label{eq:xi02dis}
\end{align}
where $a(2,4\pi T\tau)$ is defined in \eref{eq:anx}. From \eref{eq:xi02dis} and \eref{eq:chi0disPA1}, we find that $\xi_{0}$ is suppressed with disorder if 
\begin{align}
    \frac{\xi_{th}^2}{\xi_{0}^2}&> 96 \frac{a(0,4\pi T\tau)}{a(2,4\pi T\tau)},\\
    & \gtrsim 5.
\end{align}
This is always satisfied in our calculations.

\section{Detailed calculation of the self-energy from classical fluctuations in the classical renormalized regime}
\label{sec:selfcf}

Here, we detail the steps taken to obtain \eref{eq:imSelfPheno} for the imaginary part of the self-energy in the renormalized classical regime using a phenomenological approach. We start with the TPSC equation for the inelastic part of the self-energy that enters the interacting Green function
\begin{align}
    \Sigma(\mathbf{k},i\omega_n) &= \frac{U}{8}\frac{T}{N} \sum_{\mathbf{q},iq_n} \left [ 3U_{sp}\chi_{sp}(\mathbf{q},iq_n) \right. \nonumber \\
    &+ \left . U_{ch}\chi_{ch}(\mathbf{q},iq_n)\right]\mathcal{G}_0(\mathbf{k}+\mathbf{q},i\omega_n+iq_n).\label{eq:sigma2Sup}
\end{align}
We only consider the contributions to the self-energy from the spin fluctuations at $q_n=0$, which are dominant in the renormalized classical regime where the characteristic spin-fluctuation frequency is smaller than temperature
\begin{equation}
    \Sigma_{cl}(\mathbf{k},i\omega_n) \simeq \frac{3UU_{sp}}{8}\frac{T}{N} \sum_{\mathbf{q}}\chi_{sp}(\mathbf{q},0) \mathcal{G}_0(\mathbf{k}+\mathbf{q},i\omega_n).
\end{equation}
Next, we perform the analytic continuation $i\omega_n\rightarrow\omega+i\eta$ to obtain the retarded self-energy $\Sigma^R_{cl}$ and evaluate the expression at $\omega=0$. We obtain
\begin{equation}
    \Sigma^R_{cl}(\mathbf{k},\omega=0) \simeq \frac{3UU_{sp}}{8}\frac{T}{N} \sum_{\mathbf{q}}\chi_{sp}(\mathbf{q},0) \frac{1}{-\epsilon_{\mathbf{k}+\mathbf{q}}+\mu-\Sigma^{dis}},
\end{equation}
where the self-energy due to disorder takes the form $\Sigma^{dis}=-i/2\tau$ after analytical continuation. Next, we use the Ornstein-Zernicke form for the spin susceptibility which is valid in this regime of strong spin fluctuations
\begin{equation}
    \chi_{sp}(\mathbf{q},0) \simeq \frac{1}{U_{sp}\xi_0^2}\frac{2}{(\mathbf{q}-\mathbf{Q})^2 +\xi_{sp}^{-2}},
    \label{eq:chispSup}
\end{equation}
and transform the sum over $\mathbf{q}$ to an integral with the variable change $\mathbf{q}-\mathbf{Q}\rightarrow\mathbf{q}$
\begin{equation}
    \Sigma^R_{cl}(\mathbf{k},0) = \frac{3UT}{4\xi_0^2} \int \frac{d^2q}{(2\pi)^2}\frac{1}{\mathbf{q}^2+\xi_{sp}^{-2}} \frac{1}{-\epsilon_{\mathbf{k}+\mathbf{q}+\mathbf{Q}}+\mu+i/2\tau}.
\end{equation}
We evaluate at the hot spots $\mathbf{k}=\mathbf{k}_F$ where the Fermi surface crosses the AFM Brillouin zone. This means that $k_{F,y} = \pi - k_{F,x}$, which implies that $\epsilon_{\mathbf{k}_F+\mathbf{Q}}=\epsilon_{\mathbf{k}_F} = \mu$.
We also approximate $\epsilon_{\mathbf{k}_F+\mathbf{q}+\mathbf{Q}}=\epsilon_{\mathbf{k}_F+\mathbf{Q}}+\mathbf{q}\cdot\mathbf{v}_{F}$. This yields
\begin{equation}
    \Sigma^R_{cl}(\mathbf{k}_F,0) = \frac{3UT}{4\xi_0^2} \int \frac{d^2q}{(2\pi)^2}\frac{1}{q_{\perp}^2+q_{\parallel}^2+\xi_{sp}^{-2}} \frac{1}{i/2\tau-q_\parallel v_{F}},
\end{equation}
where we defined $q_{\parallel}$ ($q_\perp$) as the component of $\mathbf{q}$ that is parallel (perpendicular) to the Fermi velocity $\mathbf{v}_F$. We can then perform the integral over $q_\perp$ in the complex plane
\begin{align}
    \Sigma^R_{cl}(\mathbf{k}_F,0) &= \frac{3UT}{16\pi^2\xi_0^2} \int dq_\parallel dq_\perp \left [ \frac{1}{q_{\perp}+i\sqrt{q_{\parallel}^2+\xi_{sp}^{-2}}} \right. \nonumber \\
    & \left.- \frac{1}{q_{\perp}-i\sqrt{q_{\parallel}^2+\xi_{sp}^{-2}}} \right ] \frac{i}{2\sqrt{q_\parallel^2 +\xi_{sp}^{-2}}} \frac{1}{i/2\tau-q_\parallel v_{F}}, \nonumber\\
    &= \frac{3UT}{16\pi^2\xi_0^2} \int dq_\parallel (-2\pi i)\frac{i}{2\sqrt{q_\parallel^2 +\xi_{sp}^{-2}}} \frac{1}{i/2\tau-q_\parallel v_{F}},\nonumber \\
    &= \frac{3UT}{16\pi\xi_0^2} \int dq_\parallel \frac{1}{\sqrt{q_\parallel^2 +\xi_{sp}^{-2}}} \frac{1}{i/2\tau-q_\parallel v_{F}}.
\end{align}
Next, we focus on the imaginary part of the self-energy and perform the integral over $q_\parallel$ by first changing variables
\begin{widetext}
\begin{align}
    \mathrm{Im}\Sigma^R_{cl}(\mathbf{k}_F,0) 
    &= -\frac{3UT}{16\pi\xi_0^2} \int dq_\parallel \frac{1}{\sqrt{q_\parallel^2 +\xi_{sp}^{-2}}} \frac{1/2\tau }{(1/2\tau)^2+(q_\parallel v_{F})^2},\nonumber\\
    &= -\frac{3UT}{16\pi\xi_0^2}\frac{1}{2v_F^2 \tau} \int dq_\parallel \frac{\xi_{sp}}{\sqrt{(q_\parallel\xi_{sp})^2 +1}}  \frac{\xi_{sp}^2 }{(\xi_{sp}/2v_F\tau)^2+(q_\parallel \xi_{sp})^2},\nonumber\\
    &= -\frac{3UT}{16\pi\xi_0^2}\frac{\xi_{sp}^2}{2v_F^2 \tau} \int dx \frac{1}{\sqrt{x^2 +1}} \frac{1}{(\xi_{sp}/2v_F\tau)^2+x^2}.
\end{align}
\end{widetext}
We then integrate from $-\infty$ to $\infty$, which leads to
\begin{align}
    \mathrm{Im}\Sigma^R_{cl}(\mathbf{k}_F,0) 
    &= -\frac{3UT}{8\pi\xi_0^2}\frac{\xi_{sp}}{v_F}\frac{\mathrm{arccosh}(\xi_{sp}/2v_F\tau)}{\sqrt{(\xi_{sp}/2v_F\tau)^2-1}},\nonumber\\
    &= -\frac{3U}{8\pi^2\xi_0^2}\frac{\xi_{sp}}{\xi_{th}}\frac{\mathrm{arccosh}(\xi_{sp}/2v_F\tau)}{\sqrt{(\xi_{sp}/2v_F\tau)^2-1}},
\end{align}
where we introduced the de Broglie wavelength $\xi_{th}=v_F/\pi T$. This leads to the form used in the main text,
\begin{equation}
    \mathrm{Im}\Sigma^R_{cl}(\mathbf{k}_F,0) = -\frac{3U}{8\pi^2\xi_0^2}\frac{\xi_{sp}}{\xi_{th}}F\left(\frac{\xi_{sp}}{v_F\tau}\right),
\end{equation}
with 
\begin{align}
    F(x) =  \left \{ \begin{matrix}\frac{\pi}{2} & & x=0~\textrm{(No~disorder)},\\
        \frac{\mathrm{arccosh}(x)}{\sqrt{x^2-1}} &  & x > 0
    \end{matrix} \right.
\end{align}
where the value of $\mathrm{arccosh}$ must be computed in the complex plane. It is purely imaginary for $x < 1$.

\end{document}